\documentclass{emulateapj}
\usepackage{apjfonts}
\usepackage{graphicx}
\usepackage{amsmath}

\usepackage{color}

\newcommand{\appropto}{\mathrel{\vcenter{
  \offinterlineskip\halign{\hfil$##$\cr
    \propto\cr\noalign{\kern2pt}\sim\cr\noalign{\kern-2pt}}}}}

\def\kms{\ifmmode{\rm km\thinspace s^{-1}}\else km\thinspace s$^{-1}$\fi}

\shortauthors{Sanchis-Ojeda et al.~2013}
\shorttitle{Kepler-78b}

\begin{document}

%
\def\ltsima{$\; \buildrel < \over \sim \;$}
\def\lsim{\lower.5ex\hbox{\ltsima}}
\def\gtsima{$\; \buildrel > \over \sim \;$}
\def\gsim{\lower.5ex\hbox{\gtsima}}
%

\bibliographystyle{apj}

\title{
Transits and Occultations of an Earth-Sized Planet in an 8.5-Hour Orbit
}

\author{
Roberto~Sanchis-Ojeda\altaffilmark{1},
Saul~Rappaport\altaffilmark{1},
Joshua~N.~Winn\altaffilmark{1},
Alan~Levine\altaffilmark{2},\\
Michael~C.~Kotson\altaffilmark{3},
David~W.~Latham\altaffilmark{4}, 
Lars~A.~Buchhave\altaffilmark{5, 6}
}

\altaffiltext{1}{Department of Physics, and Kavli Institute for
  Astrophysics and Space Research, Massachusetts Institute of
  Technology, Cambridge, MA 02139, USA; rsanchis86@gmail.com, sar@mit.edu, jwinn@mit.edu}

\altaffiltext{2}{M.I.T.\ Kavli Institute for Astrophysics and Space Research, 70 Vassar St., Cambridge, MA, 02139; aml@space.mit.edu}

\altaffiltext{3}{Institute for Astronomy, University of Hawaii, 2680 Woodlawn Drive, Honolulu, HI 96822, USA}

\altaffiltext{4}{Harvard-Smithsonian Center for Astrophysics, 60 Garden St., Cambridge, MA, 02138, USA}

\altaffiltext{5}{Niels Bohr Institute, University of Copenhagen, Juliane Maries vej 30, 2100 Copenhagen, Denmark }

\altaffiltext{6}{Centre for Star \& Planet Formation, Natural History Museum of Denmark, University of Copenhagen, {\O}ster Voldgade 5-7, 1350 Copenhagen, Denmark}

 \journalinfo{{\it Astrophysical Journal}, in press}
 \slugcomment{Accepted for publication, 2013 July 11}

\begin{abstract}

  We report the discovery of an Earth-sized planet ($1.16\pm 0.19~R_\oplus$) in an 8.5-hour orbit around a late G-type star (KIC~8435766, Kepler-78). The object was identified in a search for short-period planets in the {\it Kepler} database and confirmed to be a transiting planet (as opposed to an eclipsing stellar system) through the absence of ellipsoidal light variations or substantial radial-velocity variations. The unusually short orbital period and the relative brightness of the host star ($m_{\rm Kep}$ = 11.5) enable robust detections of the changing illumination of the visible hemisphere of the planet, as well as the occultations of the planet by the star. We interpret these signals as representing a combination of reflected and reprocessed light, with the highest planet dayside temperature in the range of 2300~K to 3100~K. Follow-up spectroscopy combined with finer sampling photometric observations will further pin down the system parameters and may even yield the mass of the planet.

\end{abstract}


\section{Introduction}

The work described here was motivated by our curiosity about planets with the shortest possible orbital periods. Although many exoplanets have been discovered with orbital periods of a few days---most famously the ``hot Jupiters''---relatively few are known with periods shorter than {\it one} day. Howard et al.~(2012) found that such planets are less common than planets with periods of 2--3 days, based on data from the {\it Kepler} spacecraft. The shortest-period transit candidate is the 4.5-hour signal in the KOI~1843 system found by Ofir \& Dreizler (2012), although this candidate has not yet been thoroughly vetted. Among the well-documented planets the record holder is Kepler-42c, with a period of 10.9~hr (Muirhead et al.\ 2012). The planet 55~Cnc~e is the shortest-period planet (17.8~hr) for which the radius and mass have both been measured (Winn et al.~2011; Demory et al.~2011).  In all of these cases the planet is smaller than about 2~$R_\oplus$. Among giant planets, the shortest period belongs to WASP-19b ($P=18.9$~hr, Hebb et al.~2010).

The rarity of giant planets with $P<1$~day could reflect the vulnerability of such planets to tidally-induced decay of their orbits (see, e.g., Schlaufman et al.~2010 for specific predictions), a possible tidal-inflation instability (Gu et al.~2003a), Roche-lobe overflow (Gu et al.~2003b), or evaporation (see, e.g., Murray-Clay et al.~2009).  If so, then because smaller rocky planets are less vulnerable to these effects, one would expect smaller planets to be more common than large planets at the shortest periods. A suggestion that this is indeed the case comes from perusing the list of the active {\it Kepler} Objects of Interest (KOI)\footnote{http://exoplanetarchive.ipac.caltech.edu/}. In this list we find only 17 active planet candidates with $P<16$~hr, all of which have inferred planet sizes smaller than that of Neptune. However, this result is difficult to interpret because of the possibility of false positives due to eclipsing binary stars. There are already 55 known false positives in that period range, and the vetting is incomplete for most of the 17 candidates that remain.

There is also the possibility that the KOI catalog is missing some objects with short periods or short transit durations, as noted by Fressin et al.~(2013). Other authors have performed independent searches of the {\it Kepler} database using the Box Least Squares (BLS) algorithm (Kov{\'a}cs, Zucker, \& Mazeh 2002) and found new candidates (Ofir \& Dreizler 2012; Huang, Bakos, \& Hartman 2013). The BLS algorithm is designed for finding transit signals of short duration compared to the orbital period. For the shortest-period planets, though, the transit duration is a sizable fraction of the orbital period, and a Fourier Transform (FT) analysis should be sufficient for detection. The FT has the advantages of simplicity and speed. This was the technique used to detect the apparently disintegrating planet KIC~12557548b, with an orbital period of 15.6~hr (Rappaport et al.\ 2012).

Here we describe an Earth-sized planet with an orbital period of 8.5~hr, which was not among the {\it Kepler} Objects of Interest, and was identified in our FT-based survey of the {\it Kepler} data.  Because so many transits have been observed and the star is unusually bright (with {\it Kepler} magnitude 11.5), the process of validation is simplified, the planetary occultations and illumination variations are easily detected, and further observations should be rewarding.

Section~\ref{sec:obs} of this paper describes the initial detection of the signal, our follow-up ground-based spectroscopic observations, and the properties of the parent star. Section~\ref{sec:trans} presents the analysis of the {\it Kepler} light curve and the determination of the system parameters.  Section~\ref{sec:vali} demonstrates that the signal almost certainly arises from a transiting planet, as opposed to an eclipsing binary star, based on the lack of detectable ellipsoidal light variations or radial-velocity variations. Section~\ref{sec:energy} discusses the possibilities for the surface temperature and reflectivity of the planet, based on the observed properties of the illumination curve and occultations.  We end with a brief discussion of the future prospects for studying the shortest-period planets.

\section{Observations}
\label{sec:obs}

\subsection{Initial detection}

To carry out an independent search for the shortest-period planets, we subjected all the {\em Kepler} long-cadence light curves to a FT analysis using data from Quarters 1-14. The light curves used for this study had all been processed with the PDC-MAP algorithm (Stumpe et al.\ 2012; Smith et al.\ 2012), which removes many of the instrumental artifacts from the time series while retaining the bulk of the astrophysical variability.  The time series from each quarter was divided by its median value. Then, the normalized data from all quarters were stitched together into a single file. The FT was then calculated.  We searched for the presence of at least one peak more than 4.6 standard deviations above the local noise level in the Fourier spectrum. To be considered further, we also required that the FT exhibit at least one harmonic or subharmonic that stands out at the 3.3 $\sigma$ level. The candidates were then examined by eye. Only those that showed several harmonics with a slow falloff in amplitude with increasing frequency, and no sign of stellar pulsations, were selected for further study.

The surviving candidates underwent a period-folding analysis. To remove the slow flux variations caused by starspots and stellar rotation, we applied a moving-mean filter to the flux series, with a width in time equal to the candidate orbital period. Then we folded the time series with that period and inspected the resulting light curve, looking for the characteristic shape of a transit. We applied the same filtering algorithm to the time series of the row and column positions of the image photocenter (MOM\_CENTR) provided by the {\em Kepler} pipeline. Systems with large photocenter shifts were discarded; such large shifts indicate that the flux variations belong to a neighboring star and not the intended {\it Kepler} target. We also checked for any significant differences in the depths of the odd- and even-numbered transits, which would reveal the candidate to be an eclipsing binary with twice the nominal period. A list of 20 potentially new short-period planet candidates passed all these tests, with orbital periods between 4 and 16 hours. As expected, this list is comprised entirely of objects smaller than Neptune. We will report on the entire collection in a separate paper. For this initial report we chose to focus on KIC~8435766 (from now on designated Kepler-78b), because it has the brightest host star, one of the shortest orbital periods, and the most significant detection of the illumination curve and occultations.

\begin{figure*}[ht]
\begin{center}
\leavevmode
\hbox{
\epsfxsize=7in
\epsffile{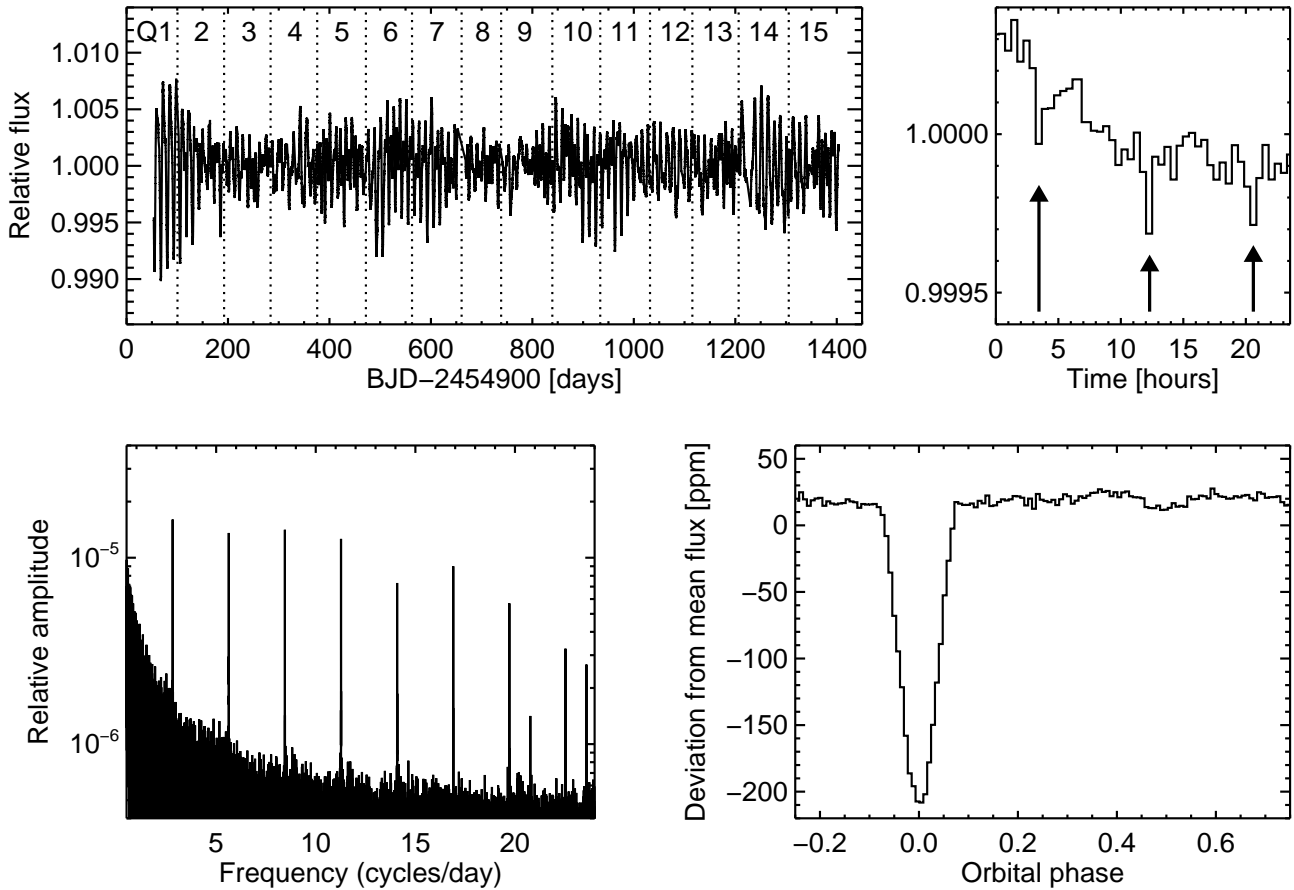}}
\end{center}
\vspace{-0.1in}
\caption{{\it Kepler} data for Kepler-78 (KIC~8435766). {\it Top.}---Time series, based on quarterly-normalized PDC-MAP data. The quasiperiodic flux variations are characteristic of starspots and rotation. The right panel focuses on one arbitrarily chosen day of data, to allow
several transits to be seen by eye (single 30 minute bins).
{\it Lower left panel.}---Fourier transform, before any filtering of the starspot-induced signal.
{\it Lower right panel.}---Light curve, after filtering and then folding the time series
with a period of 8.5201795~hr.}
\label{fig:FFT}
\vspace{0.1in}
\end{figure*}

Figure~\ref{fig:FFT} shows the time series, FT, and folded light curve for Kepler-78. The time series exhibits quasiperiodic flux variations with an amplitude of a few percent and a period of 12.5 days, likely the result of spots on a rotating star. The FT also shows a base frequency at $\nu = 3$~cycles~day$^{-1}$, and at least 9 higher harmonics, two of which are aliases resulting from reflection about the Nyquist limit of 25~cycles~day$^{-1}$.  No subharmonics of these frequencies are seen; a positive detection would have been indicative of a binary with primary and secondary eclipses of nearly equal depth. The folded light curve shows a transit with a depth of 220~ppm and an occultation with a depth of 10~ppm. The illumination curve---the rise in flux between transit and occultation---is less obvious; see Figure~\ref{fig:fulmodel} for a better view.

\subsection{Spectroscopy}
\label{sec:spectroscopy}

Spectroscopic observations were undertaken to characterize the host star and to search for radial-velocity variations. We obtained five spectra with the fiber-fed Tillinghast Reflector {\'E}chelle Spectrograph on the 1.5m Tillinghast Reflector at the Fred Lawrence Whipple Observatory on Mt.\ Hopkins, Arizona. The observations took place in 2013 on March 23, 25, and 29, and on April 2 and 4. Individual exposure times of about 15 minutes yielded a signal-to-noise ratio (S/N) per resolution element in the Mg~I~b order ranging from 26 to 44, depending mostly on the seeing and sky transparency.

Spectroscopic parameters for the host star were determined with the Stellar Parameter Classification code (SPC; Buchhave et al.~2012). We derived the parameters from each spectrum, and then computed weighted averages. The results were $T_{\rm eff} =5089 \pm 50$~K, ${\rm [m/H]} = -0.14 \pm 0.08$, $\log g = 4.60 \pm 0.1$, and $v \sin i = 2.4 \pm 0.5~\kms$. We also estimated the radial velocity of the star at $\gamma = -3.44 \pm 0.08~\kms$.

To estimate the stellar mass and radius, we used the calibrated relationships provided by Torres \& Andersen (2010) between the spectroscopic parameters and stellar dimensions. These relationships give a stellar mass $0.81 \pm 0.05 \,M_\odot$, radius $0.74^{+0.10}_{-0.08} \,R_\odot$, and mean density $\langle \rho \rangle = 2.8^{+1.1}_{-0.8}$ g~cm$^{-3}$.  Error propagation was performed assuming independent Gaussian errors in the stellar parameters, along with a 6\% systematic error in the stellar mass due to the uncertainty in the calibration formulas. The final values are summarized in Table~\ref{tbl:params}. Following Torres et al.\ (2012) we checked the derived mass and the radius by comparing the spectroscopic parameters to the outputs of stellar evolutionary models (Yi et al.~2001); the resulting values were similar.

We can also estimate the age of the star based on the rotation period.  Schlaufman (2010) gave a calibrated polynomial formula relating stellar age, mass, and rotation period.  Given the preceding estimates for the mass and rotation period, the Schlaufman (2010) formula gives an age of $750\pm 150$~Myr.  As a consistency check we note that the stellar radius of $0.74~R_\odot$ and the rotation period of 12.5 days imply a rotation velocity of 3~km~s$^{-1}$, which is compatible with the spectroscopic estimate of $v\sin i$ (assuming $\sin i \approx 1$).

Radial velocities were determined via cross-correlation against synthetic templates, as described by Buchhave et al.\ (2010) and Quinn et al.\ (2012).  The results are presented in Table~\ref{tab:radtab}, where the mean radial velocity has been subtracted. The five data points have a standard deviation of 32~m~s$^{-1}$, and internally-estimated measurement uncertainties of 20-25~m~s$^{-1}$.  By fitting a sinusoid with the same period and phase as the transit signal, we obtain a velocity semiamplitude $K=36\pm 12$~m~s$^{-1}$. However, the true uncertainty is undoubtedly larger because of the spurious radial velocities produced by rotating starspots, which are expected to be of order 30~m~s$^{-1}$ (the product of $v\sin i$ and the 1\% photometric modulation). Hence we consider the data to be consistent with no radial-velocity variation. Any variation with the same period and phase as the transit signal is smaller than about 100~m~s$^{-1}$ (3$\sigma$), corresponding to a companion mass of about 0.3~$M_{\rm Jup}$ (100~$M_\oplus$) orbiting Kepler-78.

\begin{deluxetable}{crr}
\tabletypesize{\scriptsize}
\tablecaption{Relative Radial Velocities for Kepler-78\label{tab:radtab}}
\tablewidth{0pt}

\tablehead{
\colhead{BJD-2456300} & \colhead{RV (m~s$^{-1}$)} & \colhead{RV$_e$\tablenotemark{a} (m~s$^{-1}$)} 
}

\startdata
74.9484 & 23.66 & 22.62 \\
76.9663 & -36.37 & 19.13 \\
81.0154 & 19.65 & 20.21 \\
84.9219 & 26.03 & 19.13 \\
86.9066 & -32.99 & 25.51
\enddata
\tablenotetext{a}{Estimated 1-$\sigma$ uncertainty in the relative radial velocity.}

\end{deluxetable}

\subsection{UKIRT image}

The {\it Kepler} time series is based on summing the flux within an aperture surrounding the target star Kepler-78 specific to each ``season'' (the quarterly 90$^\circ$ rotations of the field of view). The aperture dimensions change with the season; they are as large as 5 pixels (20$''$) in the column direction and 6 pixels (24$''$) in the row direction, with a total of 12-20 pixels used in a given season. To check whether the summed flux includes significant contributions from known neighboring stars, we examined the $J$ band image from the UKIRT survey of the {\it Kepler} field\footnote{http://keplergo.arc.nasa.gov/ToolsUKIRT.shtml}.

Two neighboring stars were detected (see Figure~\ref{fig:centroids}). Relative to Kepler-78, the first neighbor is 4.5~mag fainter and 4.8$''$ away, contributing 1.5\% of the $J$-band flux to the {\it Kepler} photometric aperture.  The second neighbor is 3.1~mag fainter and 10.3$''$ away. If this star were wholly within the aperture it would have contributed 5.5\% of the total $J$-band flux; however, since it falls near the edge of the aperture, the contribution is likely smaller and is expected to vary with the {\it Kepler} seasons. In section~\ref{sec:vali} we will show
that neither of these fainter stars can be the source of the transit signal.

\section{Light curve analysis}
\label{sec:trans}

\begin{deluxetable*}{lcc}
\tabletypesize{\scriptsize}
\tablecaption{System Parameters of Kepler-78\lowercase{b}\label{tbl:params}}
\tablewidth{0pt}

\tablehead{
\colhead{Parameter} & \colhead{Value} & \colhead{68.3\% Conf.~Limits} 
}

\startdata
KIC number & 8435766 & ... \\
R.A.\ (J2000) & 19h 34m 58.00s & ... \\
Decl.\ (J2000) & 44\farcs26\farcm53\farcs99s & ... \\

 & & \\

Effective temperature, $T_{\textrm{eff}}$~[K]\tablenotemark{a}       &   5089  & $\pm 50$ \\  
 
Surface gravity, $\log g$~[$g$ in cm s$^{-2}$]\tablenotemark{a}  & 4.60 & $\pm 0.1$ \\   
 
Metallicity, [m/H]\tablenotemark{a}  & $-0.14$ & $\pm 0.08$ \\

Projected rotational velocity, $v \sin i$~[km~s$^{-1}$]\tablenotemark{a}    & 2.4 & $\pm 0.5$ \\

Radial velocity of the star, $\gamma$~[km~s$^{-1}$] & $-3.44$ & $\pm 0.08$ \\

Mass of the star, $M_\star$~[$M_{\odot}$]\tablenotemark{b}                  & $0.81$    & $\pm 0.05$  \\

Radius of the star, $R_\star$~[$R_{\odot}$]\tablenotemark{b}        &  $0.74$  &  $+0.10$ $-0.08$ \\

``Spectroscopic'' mean stellar density, $\langle \rho_\star\rangle $~[g~cm$^{-3}$]\tablenotemark{b} &  $2.8$  & $+1.1$ $-0.8$ \\

Rotation period~[days] & 12.5 & $\pm 1.0$ \\ 

 & & \\

Reference epoch~[BJD$_{\rm TDB}$]                      & $2454953.95995$  &  $\pm 0.00015$      \\

Orbital period~[days]                                 & $0.35500744$      &  $\pm 0.00000006$    \\

Square of planet-to-star radius ratio, $(R_{\textrm{p}}/R_\star)^2 $ [ppm]          & $201$         &  $+57$, $-18$  \\

Planet-to-star radius ratio, $R_{\textrm{p}}/R_\star $        & $0.0142$         &  $+0.0019$, $-0.0007$  \\
Scaled semimajor axis, $a/R_\star$                    & 3.0 & $+0.5, -1.0$  \\
Orbital inclination, $i$~[deg]                        & $79$           &  $+9, -14$    \\
Transit duration (first to fourth contact)~[hr]                 & $0.814$   &   $+0.021$,  $-0.015$    \\
``Photometric'' mean stellar density\tablenotemark{c}, $\langle \rho_\star \rangle$~[g~cm$^{-3}$]   &  $3.8$  & $+2.2, -2.7$   \\

 & & \\
Seasonal dilution parameters~[\%]\tablenotemark{d}  & 3.5, 0, 0.9, 5.5 & $\pm$ 1.2, 0, 1.2, 1.2 \\
Occultation depth, $\delta_{\rm occ}$ [ppm] & 10.5 & $\pm 1.2$ \\
Amplitude of illumination curve, $A_{\rm ill}$ [ppm] & 4.4 & $\pm 0.5$ \\
Amplitude of ellipsoidal light variations, $A_{\rm ELV}$ [ppm] & $<$1.2 &  (3$\sigma$) \\ 
\\

Planet radius, $R_{\textrm{p}}$~[R$_\oplus$]           &  $1.16$         &  $+0.19, -0.14$ \\
Planet mass, $M_{\textrm{p}}$~[$M_\oplus$]\tablenotemark{e}          & $<$8   &  (3$\sigma$)

\enddata

\tablecomments{The {\it Kepler} input catalog gives magnitudes $m_{\rm Kep} =11.55$,
$g=12.18$, $r=11.46$, $J=10.18$, and $T_{\rm eff}=4957\pm 200$~K.
}
\tablenotetext{a}{Obtained from an SPC analysis of the spectra.}
\tablenotetext{b}{Based on the relationships from Torres \& Andersen (2010).}
\tablenotetext{c}{Defined as $\langle \rho_\star \rangle = (3\pi/GP^2)(a/R_\star)^3$}
\tablenotetext{d}{In order, these refer to season 0, 1, 2 and 3.}
\tablenotetext{e}{Based on absence of ellipsoidal light variations, assuming zero dilution.}

\end{deluxetable*}

\subsection{Transit times and orbital period}
\label{subsec:timing}

In the first step of the light-curve analysis we determined the orbital period $P$, and checked for any transit-timing variations (TTV). Using the initial FT-based estimate of $P$, we selected a 6-hour interval (12 data points) surrounding each predicted transit time.  We corrected each transit interval for the starspot-induced flux modulation by masking out the transit data (the central 2 hours), fitting a linear function of time to the out-of-transit data, and then dividing the data from the entire interval by the best-fitting linear function.  A total of 3378 individual transits were detected and filtered in this way; a few others were detected but not analyzed further because fewer than 2 hours of out-of-transit data are available.

\begin{figure}[ht]
\begin{center}

\leavevmode
\hbox{
\epsfxsize=3.6in
\epsffile{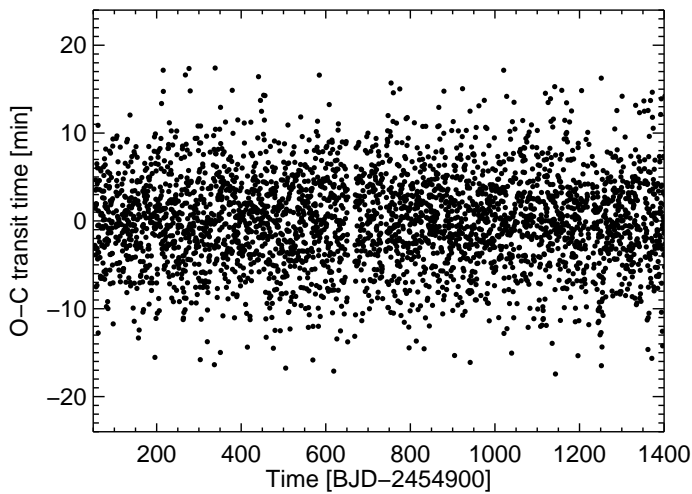}}
\end{center}
\vspace{-0.1in}
\caption{Deviations of individual transit times from strict periodicity.}
\label{fig:TTV}
\vspace{0.1in}
\end{figure}

In order to obtain precise transit times, we first need to obtain an empirical transit template. For that, the time series was folded with the trial period. As seen in Figure~\ref{fig:FFT}, the transit shape can be approximated by a triangular dip, due to the 30-minute averaging time of the {\it Kepler} observations. For transit timing purposes we used a triangular model, with three parameters describing the depth, duration, and time of the transit. We found the best-fitting model to the phase-folded light curve, and then fitted each individual transit with the same model, allowing only the time of transit to be a free parameter. As an estimate of the uncertainty in each data point, we used the standard deviation of the residuals of the phase-folded light curve.

The mean orbital period $P$ and a fiducial transit time $T_c$ were determined from a linear fit to the individual transit times, after performing 5 $\sigma$ clipping to remove outliers. Table~\ref{tbl:params} gives the results, and Figure~\ref{fig:TTV} shows the $O-C$ residuals.  We searched the residuals for periodicities in the range 10-1000~days using a Lomb-Scargle periodogram (Scargle 1982), but found none with false alarm probability lower than 1\%.

To search for any secular variation in the period, such as a period decrease due to tidal decay, we tried modeling the transit times with a quadratic function. The fit did not improve significantly. Based on this fit, the period derivative must be $| dP/dt| < 3.5 \times \,10^{-10}$ $(2\sigma)$. Using Kepler's third law, this can be transformed into a lower bound on the tidal decay timescale $ a/\dot{a}$ of 4~Myr. This is not a particularly interesting bound, given that the stellar age is estimated to be 750~Myr.

\subsection{Transit and illumination curve analysis}
\label{subsec:timeseries}

\begin{figure*}[ht]
\begin{center}
\leavevmode
\hbox{
\epsfxsize=5.4in
\epsffile{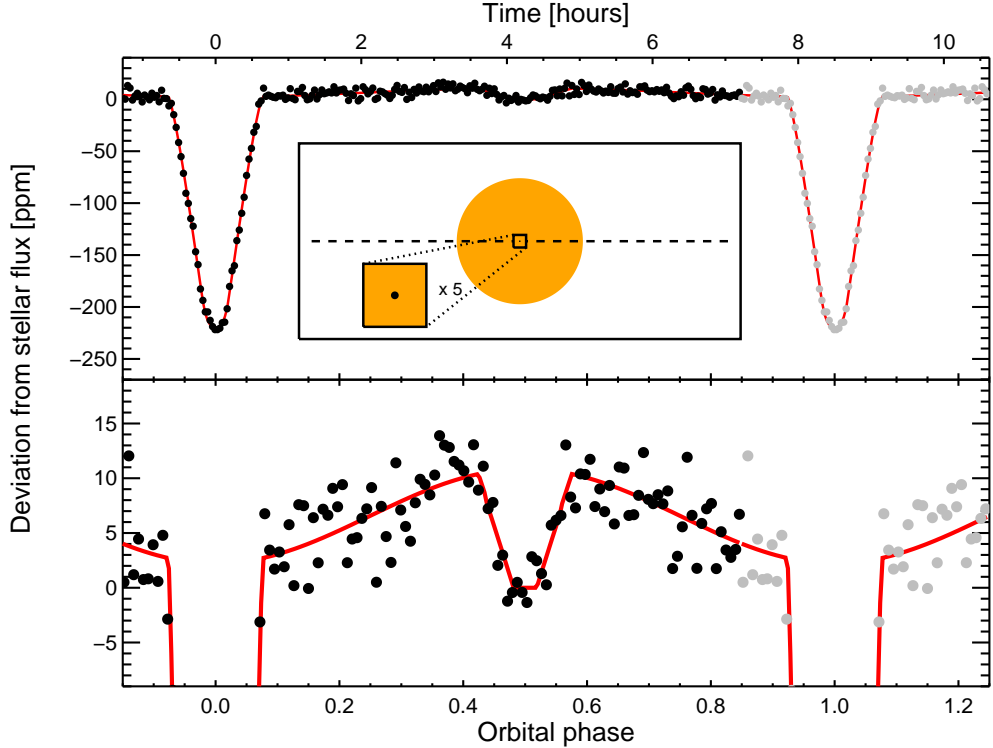}}
\end{center}
\vspace{-0.1in}
\caption{ {\it Upper panel.}---The final light curve (dots)
and best-fitting model (red curve). The inset panel is a scale illustration of the system, where the length of the dashed line represents one possible orbital distance (both the impact parameter and the orbital distance are highly uncertain). The transits look V-shaped because of the 30 minute sampling, but shorter-cadence observations should reveal a much longer flat bottom in the middle of the transit.
{\it Lower panel.}---Close-up of the illumination curve and occultation.
The data have been binned to 4~min for clearer visual inspection. }
\label{fig:fulmodel}
\vspace{0.1in}
\end{figure*}

We then returned to the original time series and repeated the process of filtering out the starspot-induced variations, this time with the refined orbital period $P$ and a slightly different procedure. The basic idea was to filter out any variability on timescales longer than the orbital period. First, we divide the flux series on each quarter by its mean. Then, for each data point $f(t)$, a linear function of time was fit to all the out-of-transit data points at times $t_j$ satisfying $|t-t_j| < P/2$. Then $f(t)$ was replaced by $f(t) -f_{\rm fit}(t) +1$, where $f_{\rm fit}$ is the best-fitting linear function.  Figure~3 shows the resulting light curve, after further correcting for seasonal-specific dilution as described below.

Subsequent analysis was restricted to data from quarters 2-15, since quarters 0 and 1 had shorter durations and the data seem to have suffered more from systematic effects.  For each season, we phase-folded the data with period $P$ and then reduced the data volume by averaging into 2~min samples. We then fitted a model including a transit, an occultation, and orbital phase modulations.

The transit model $f_{\rm tran}(t)$ was based on the Mandel \& Agol (2002) equations for the case of quadratic limb darkening. The parameters were the midtransit time $t_0$, the zero-limb-darkening transit depth $\delta_{\rm tran} = (R_p/R_\star)^2$, the impact parameter $b$, the scaled orbital separation $a/R_{\star}$, and the limb darkening coefficients $u_1$ and $u_2$.  The orbital period was held fixed, and a circular orbit was assumed.  The limb-darkening coefficients were allowed to vary, but the difference $u_1 - u_2$ was held fixed at 0.4, and the sum $u_1+u_2$ was subjected to a Gaussian prior of mean 0.7 and standard deviation 0.1. These numerical values are based on the theoretical coefficients given by Claret \& Bloemen (2011).

The occultation model $f_{\rm occ}(t)$ was a simple trapezoidal dip, centered at an orbital phase of 0.5, and with a total duration and ingress/egress durations set equal to those of the transit model. The only free parameter was the depth $\delta_{\rm occ}$.

The out-of-transit modulations were modeled as sinusoids with phases and periods appropriate for ellipsoidal light variations (ELV), Doppler boosting (DB), and illumination effects (representing both reflected and reprocessed stellar radiation). Expressed in terms of orbital phase $\phi = (t-t_c)/P$,
these components are
\begin{equation}
A_{\rm DB}\sin(2\pi\phi)- A_{\rm ELV}\cos(4\pi\phi)  - A_{\rm ill}\cos(2\pi\phi)
\label{eq:phase}
\end{equation}

A constant was added to the model flux, specific to each of the 4 seasons, representing light from neighboring stars. Since a degeneracy prevents all 4 constants from being free parameters, we set this ``dilution flux'' equal to zero for season 1, for which initial fits showed that the dilution was smallest.  The other constants should be regarded as season-specific differences in dilution.  The final parameter in the model is an overall flux multiplier, since only the relative flux values are significant.  In the plots to follow, the normalization of the model and the data was chosen such that the flux is unity during the occultations, when only the star is seen.

For comparison with the data, the model was evaluated once per minute, and the resulting values were averaged in 29.4-min bins to match the time averaging of the {\it Kepler} data.  Initial fits showed that the ELV and DB terms were consistent with zero. The non-detection of the DB term can, in principle, also be used to place upper bounds on several other effects, like inhomogeneities of the planetary albedo or a displacement of the hottest atmospheric spot on the surface of the planet with respect to the substellar point (Faigler et al. 2013).  In the absence of these phenomena,  the DB signal is expected to be negligible for this system ($\sim 0.02$~ppm), so we set $A_{\rm DB}=0$ in subsequent fits.   We optimized the model parameters by minimizing the standard $\chi^2$ function, and then used the best-fitting dilution parameters to correct all of the data to zero dilution.

Finally, we combined all of the dilution-corrected data to make a single light curve with 2-min sampling, and determined the allowed ranges for the model parameters using a Monte Carlo Markov Chain algorithm. The uncertainties in the flux data points were assumed to be identical and Gaussian, with magnitude set by the condition $\chi^2=N_{\rm dof}$. Figure~\ref{fig:fulmodel} shows the final light curve, with 4-min sampling, and the best-fitting model. Table~\ref{tbl:params} gives the results for the model parameters.

Some of these transit parameters might be affected by our choice for the filter. Using a filtering interval length of 2$P$ or 3$P$ rather than $P$ gives similar-looking light curves but with increased scatter, as expected, since the accuracy of the linear approximation for the stellar flux modulation degrades for longer time intervals. We fit these two noiser light curves with the same model and found that most of the transit parameters are not changed significantly. The secondary eclipse depth obtained were 9.8 and 10.0 ppm, slightly smaller than the value quoted on Table~\ref{tbl:params}. In the case of $A_{\rm ill}$, we obtained 4.30 and 4.65 ppm.  We set the systematic error induced by the filtering to be equal to the standard deviation of the three values obtained with the three different filtering periods, and add those in quadrature to the uncertainties obtained from the MCMC routine.  In both cases this procedure increased the uncertainties by only 10\%.  

One notable result is that the transit impact parameter is nearly unconstrained, i.e., $0 \leq b \leq 0.9$.  This is because the ingress and egress duration are poorly constrained, due to the 30~min averaging of the {\it Kepler} long-cadence data. For the same reason, $a/R_\star$ is poorly determined.

We were able to place an upper limit on the ELV amplitude of $<$1.2~ppm (3$\sigma$).
This can be translated into an upper bound on the mass of the transiting companion
using the formula (Morris \& Naftilan 1993; Barclay et al.~2012):
\begin{eqnarray}
A_{\rm ELV} & = & \left[\frac{0.15(15+u)(1+g)}{3-u} \right]\frac{M_{\rm p}}{M_{\star}} \left( \frac{R_{\star}}{a} \right)^3 \sin^2 i \nonumber \\
& \approx & 1.5 \,\, \frac{M_{\rm p}}{M_{\star}} \left( \frac{R_{\star}}{a} \right)^3 
\label{eq:ELV}
\end{eqnarray}
where $u$ is the linear limb-darkening coefficient
and $g$ is the gravity-darkening coefficient. In this case we expect
$u\approx 0.65$ and $g\approx 0.5$ (Claret \& Bloemen 2011). The upper limit on the ELV amplitude
thereby corresponds to a mass limit $M_p < 8~M_\oplus$ (3$\sigma$). (This assumes
the photometric signal is entirely from the transited star, with no dilution from
neighboring stars; see the following section for a discussion of possible dilution.)

The preceding analysis assumed the orbit to be circular, which is reasonable given the short orbital period and consequently rapid rate expected for tidal circularization.  We can also obtain empirical constraints on the orbital eccentricity $e$ based on the timing and duration of the occultation relative to the transit.  For this purpose we refitted the data adding two additional parameters, for the duration and phase of the occultation.  As expected the data are compatible with a circular orbit, and give upper limits $\left| e\cos\omega \right| < 0.016$, $\left| e\sin\omega \right| < 0.15$ and $e<0.24$ (3$\sigma$).

\section{Validation as a planet}
\label{sec:vali}

What appears to be the shallow transit of a planet may actually be the eclipse of a binary star system superposed on the nonvarying light from the brighter intended target star. The binary could be an unrelated background object, or it could be gravitationally bound to the target star and thereby be part of a triple star system. In this section we examine these possibilities, and find it more likely that the signal actually arises from the transits of a planet.
 
\subsection{Image photocenter motion}

We start with an analysis of the image photocenter location, to try to extract information about the spatial location of the varying light source (see for example Jenkins et al.\ 2010). For each of the four {\em Kepler} seasons, we filtered the time series of the photocenter row and column pixel coordinates in the same manner that was described in Section~\ref{subsec:timeseries} for the flux time series. For each of these time series, we calculated the mean of the in-transit coordinate, the mean of the out-of-transit coordinate, and the differences between those means, which we denote $dx$ and $dy$.  Using the filtered flux time series for each season, we also calculated the mean of the in-transit fluxes and the mean of the out-of-transit fluxes, both normalized to the overall mean flux. We denote the difference between these two means as $df$.

\begin{figure}[ht]
\begin{center}
\leavevmode
\hbox{
\epsfxsize=3.3in
\epsffile{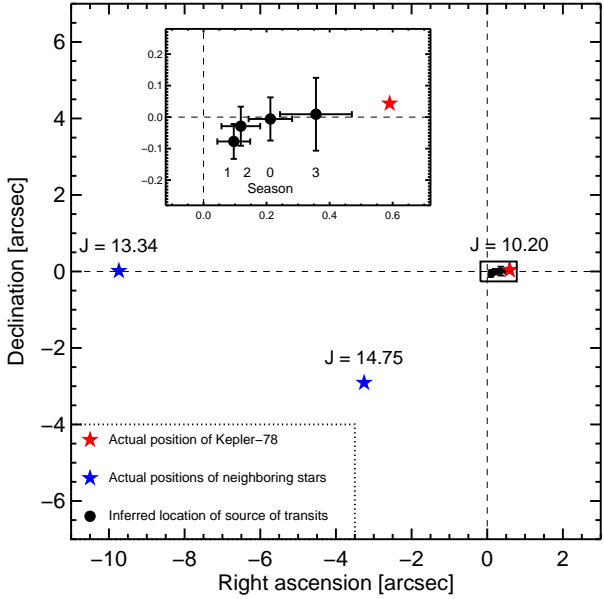}}
\end{center}
\vspace{-0.1in}
\caption{ Position of the three brightest stars (star symbols) in the vicinity of Kepler-78 relative to the center of light in J band (intersection of dashed lines). The positions of the source of the transits obtained from the centroid shifts (see text), represented by solid small circles with error bars, are close to the position of Kepler-78. The inset shows an expanded view of the vicinity of Kepler-78. }
\label{fig:centroids}
\vspace{0.1in}
\end{figure}

We then examined the ratios $dx/df$ and $dy/df$. When either of these is multiplied by the pixel size (4$\arcsec$), one obtains the angular offset between the varying source of light and the out-of-transit image photocenter. One can then compare these offsets to the locations of the stars revealed in the UKIRT images. Since the celestial coordinates of the signals in the {\em Kepler} aperture are difficult to obtain with high accuracy, we used the center-of-light of the three detected stars in the $J$-band UKIRT image as our estimate for the celestial coordinates of the out-of-transit photocenter of the {\it Kepler} signal.  The resulting determinations of the spatial location of the varying light source are shown in Figure~\ref{fig:centroids}. These results indicate that the source of the transit signal cannot be either of the two known neighboring stars. The variable source must be located within $1\arcsec$ of Kepler-78,  close enough to severely restrict the possibility of a blend with a background binary (Morton \& Johnson 2011).

The inferred coordinates of the time-variable source are correlated with the {\em Kepler} seasonal dilution parameters (see Table~\ref{tbl:params}), in a manner that is consistent with variable contamination by the brighter
of the two neighboring stars. In season 1, when the diluting flux was found to be smallest, the centroid shift was also smallest ($0.10 \pm 0.06$ arcsec). A zero centroid shift during transit implies that other sources of light have a negligible contribution to the total light within the aperture.  Based on this, the brighter neighboring star can contribute no more than 1-2\% of the total flux within the aperture during season 1.  This supports our choice of a dilution parameter equal to zero for season 1.

This analysis of the photocenter motion is sufficient for our purposes.  However, as pointed out by Bryson et al.~(2013), for a more detailed analysis it would be advisable to fit the pixel data using the {\em Kepler} Pixel Response Function, rather than using moment-based centroids as we have done.

\subsection{Ellipsoidal light variations}

The photocenter analysis cannot rule out the possibility that the observed phenomena are due to a background eclipsing binary gravitationally bound to the target star in a triple system or coincidentally within $1\arcsec$ of the target star.  In such cases, though, the ELVs would generally be larger than the observed upper bound, as we show presently. This technique was
recently used by Quintana et al.~(2013) to validate a hot Jupiter.

In Section~\ref{subsec:timeseries} we found that the upper limit on $A_{\rm ELV}$ leads to an upper limit on the companion mass of $8~M_\oplus$. This analysis was based on the assumption that the dilution was small. If the signal arises from a faint unresolved eclipsing binary, the ELV signal could have been diluted by a large factor, weakening the constraint on the companion mass. However, an upper bound on the dilution factor can be obtained from further analysis of the folded light curve. This can be understood qualitatively as follows.  If the dilution is severe, then the true transit/eclipse is much deeper than 200~ppm and the ratio of the radii of the secondary
and primary star must be larger than the ratio inferred assuming no dilution. A large radius ratio implies relatively long ingress and egress durations, which at some point become incompatible with the observed light curve shape. The sensitivity of this test is hindered by the 29.4~min cadence of the {\it Kepler} data, but in this case it still provides a useful constraint.

For a quantitative analysis we reanalyzed the phase-folded light curve in a manner similar to that presented in Section~\ref{subsec:timeseries}.  We suppose that the observed flux is $f(t) = f_0(t) + C$, where $f_0$ is the undiluted transit/occultation signal and $C$ is the diluting flux. Since our model is always normalized to have unit flux during the occultation, we use units in which $f_0(t)=1$ during the occultation, and we also defined a normalized version of $f(t)$,
\begin{equation}
f_n(t) = \frac{f_0 + C}{1 + C}.
\end{equation}
With this definition $A_{\rm trans} \approx  (1+C) A_{\rm trans,0}$,
where $A_{\rm trans,0}$ is the observed depth of the transit and $A_{\rm trans}$ is the actual depth before dilution. We stepped through a range of values for $C$, finding the best-fitting model in each case and recording the goodness-of-fit $\chi^2$. We found $C<370$ with 3$\sigma$ confidence. 

In the presence of dilution, Eqn.~(\ref{eq:ELV}) can be used to solve for the
mass ratio,
\begin{equation}
\label{eq:elv-dilution}
\frac{m}{M} \approx \frac{2}{3}~(1+C)~A_{\rm ELV} \left( \frac{a}{R} \right)^3,
\end{equation}
using notation that does not presume the secondary is a planet
($m$ and $M$ are the secondary and primary masses, and $R$ is the primary radius).
An upper limit on the mass ratio follows
from the constraints $C<370$, $A_{\rm ELV} < 1.2$~ppm,
and $a/R \lesssim 1.25 P/(\pi T)$. The latter constraint follows
from the requirement that the maximum eclipse duration
of $\approx$$1.25RP/\pi a$ exceed the observed eclipse
duration $T$; the factor of 1.25 arises from the fact that at the maximum
possible dilution the secondary star is approximately 1/4 the size of the primary
star. The result is 
\begin{equation}
\frac{m}{M} \lesssim \frac{2}{3}~C~A_{\rm ELV} \left( \frac{1.25P}{\pi T} \right)^3 \approx 0.023 ~~~.
\label{eq:limit1}
\end{equation}

Having established that $m\ll M$ regardless of dilution,
we may obtain a second condition on the mass ratio
by using Kepler's third law, $GM/a^3 = (2\pi/P)^2$,
to eliminate $a$ from Eqn.~(\ref{eq:elv-dilution}). This gives
\begin{equation}
\frac{m}{M} \approx \frac{2}{3} C~A_{\rm ELV} \left(\frac{GM}{R^3}\right)
\left( \frac{P}{2\pi} \right)^2,
\end{equation}
which leads to another inequality,
\begin{equation}
m < 3~M_{\rm Jup}
\left(\frac{M}{M_\odot}\right)^2 
\left(\frac{R}{R_\odot}\right)^{-3}.
\end{equation}
The scaling with $R^{-3}$ makes clear that the most massive
secondaries are allowed for small primary stars, i.e., main-sequence
stars and not giants. On the lower main sequence, $R\appropto M$, allowing
this inequality to be expressed purely as a function of primary mass,
\begin{equation}
\label{eq:condition}
m < 3~M_{\rm Jup}
\left(\frac{M}{M_\odot}\right)^{-1}.
\end{equation}
The maximum value of $m$ that satisfies both
Eqn.~(\ref{eq:condition}) and Eqn.~(\ref{eq:limit1}) is $8~M_{\rm Jup}$, for $M=0.36~M_\odot$.
On the upper main sequence, $R\appropto M^{0.6}$ and there is little variation
in $M^2/R^3$. Hence regardless of the primary mass, the secondary
mass is in the planetary regime.

These ELV-based constraints allow for the possibility that the secondary is a giant planet or ``super-Jupiter.''  However, we find scenarios involving a giant planet to be implausible. This is because no giant planet with a period $<$10~hr has ever been detected by any transit or Doppler survey, despite those surveys' strong sensitivity to such objects; it would be peculiar indeed for the first such system to be discovered as a blend with a {\it Kepler} target star.  This is in addition to the theoretical problems of tidal decay, tidal inflation instability, Roche-lobe overflow, and evaporation that were mentioned in Section 1. Finally the absence of detectable image photocenter motion significantly reduces the probability of blend scenarios. We henceforth assume that the dilution is small and that the system is what it appears to be at face value: an Earth-sized planet in an extremely tight orbit around the target star.

\section{Simple Physical Models}
\label{sec:energy}

It is unusual to have access to the occultation and illumination signals of such a small planet. The only other terrestrial-sized planet for which these signals have been clearly detected is Kepler-10b (Batalha et al.~2011), for which the planet is larger and the signal-to-noise ratio is lower than for Kepler-78b.

Extremely hot rocky planets are expected to be tidally-locked and have low-pressure atmospheres, free of volatiles which are removed from their surfaces by high intensity stellar winds and extreme-UV fluxes. Their atmospheres should mostly be composed of heavier-element vapors with low pressure, which will not be efficient at bringing heat from the dayside to the nightside of the planet.  L{\'e}ger et al.~(2011) have modeled many of these effects, and proposed to call such objects ``lava-ocean planets.''  If the nightside of the planet is much darker than the dayside we expect $\delta_{\rm occ} \approx 2A_{\rm ill}$, i.e., the system flux should be about the same during occultations as it is when only the nightside of the planet is in view. This has also been observed for some hot Jupiters, especially those with the highest temperatures, a sign that it is difficult to circulate heat to the night side (Cowan \& Agol 2011). Our light-curve analysis gives $\delta_{\rm occ}/2A_{\rm ill} = 1.20 \pm 0.15$, compatible with unity within 1.3$\sigma$. This is a good consistency check on the interpretation as a planet. If we repeat the light-curve analysis after imposing the constraint $\delta_{\rm occ} = 2A_{\rm ill}$, we find $\delta_{\rm occ} = 9.2 \pm 0.8$ ppm vs.~the value of $10.5 \pm 1.2$ ppm without this constraint.

In this section we use simple physical models to interpret the occultation and illumination signals. These models assume that there is no mechanism for redistributing the energy across the planetary surface. Reflection, absorption and reradiation occur locally.
Each element of the planet surface reflects a portion of the incident starlight and absorbs the remainder. The spectrum of the reflected starlight is taken to be identical to that of the incident starlight. The absorbed starlight heats the surface, which then radiates with a blackbody spectrum characterized by the local temperature. The directionality of both the reflected and thermally-emitted radiation are assumed
to follow Lambert's law. Under these assumptions, the geometric albedo is 2/3 of the Bond albedo (Cowan \& Agol 2011).

The equilibrium temperature $T_{\rm eq}$ of each element of the planet's surface is
found through the expression
\begin{equation}
\sigma_{\rm SB} T_{\rm eq}^4 = (1 - A)F_{\rm inc},
\end{equation}
where $A$ is the Bond albedo, $F_{\rm inc}$ is the power per unit surface
area of the incident starlight, and $\sigma_{\rm SB}$ is the
Stefan-Boltzmann constant.  The finite size of the star and its finite
distance from the planet are taken into account in computing $F_{\rm
inc}$ via the evaluation of a numerical integral. The equilibrium
temperature varies from a maximum value at the substellar
point to zero in those regions where no part of the star is visible.

The {\em Kepler} telescope and detectors respond differently to
radiation sources with different spectra. The differences are
incorporated into our simple models via ratios of bolometric
correction factors. These bolometric corrections were
calculated with an accuracy adequate
for the present purposes by numerically integrating blackbody spectra
over the bandpass of the {\it Kepler} observations.\footnote{http://keplergo.arc.nasa.gov/CalibrationResponse.shtml}
The stellar spectrum was taken to be that of a 5089~K blackbody.

Models were made for each of three sets of system parameters. The crucial parameters $\{a/R_\star, R_{\rm p}/R_\star\}$ took on the values $\{2.5, 0.0160\}$, $\{2.9, 0.0145\}$ and $\{3.3, 0.0130\}$ in the three models, spanning the reasonable range for these two parameters. 
The orbital inclination was set to give the observed transit duration. For each of the three models, the
calculations were carried out for values of the Bond albedo $A$
ranging from 0 to 1 in steps of 0.02.

\begin{figure}[ht]
\begin{center}

\leavevmode
\hbox{
\epsfxsize=3.4in
\epsffile{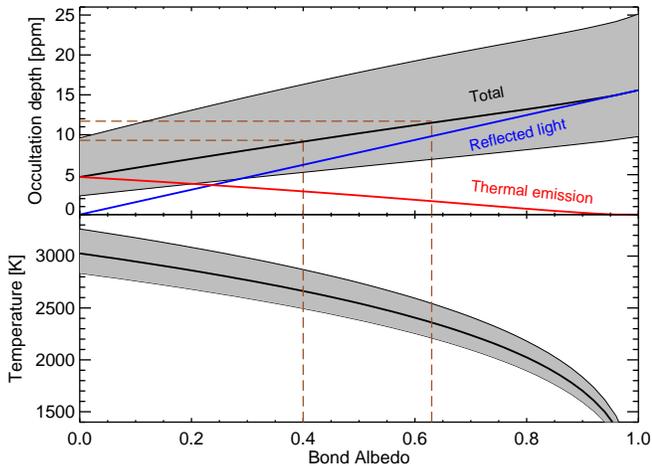}}
\end{center}
\vspace{-0.1in}
\caption{{\it Upper panel.}---Calculated occultation depth $\delta_{\rm occ}$ in parts per
million as a function of the planet albedo.  For the best fit solution the red curve shows the contribution from the thermal emission, while the blue curve represents the reflected light component.  The brown dashed horizontal lines show the $\pm1\ \sigma$ region of the measured depth, and the vertical lines represent the allowed albedo values. {\it Lower panel.}--- Maximum planet surface temperature as a function of albedo.  In both panels, the shaded regions represent the $\pm$ 1 sigma interval of possible values. }
\label{fig:albedo}
\vspace{0.1in}
\end{figure}

The results of the calculations are shown in Figure~\ref{fig:albedo}.  All three models yielded occultation depths consistent with the measured occultation depth (within 1$\sigma$). The thick black line represents the best-fitting models, which favor $A=0.4$--0.6.  However the uncertainty in the transit parameters remains large enough that any value of the albedo is allowed.  In all of these models, the occultation depth is very nearly equal to the peak-to-peak amplitude of the illumination curve, as observed.  Figure~\ref{fig:albedo} also shows the model-derived maximum planet surface temperatures, which occur at the substellar point.  The possibilities range from $T_{\rm eq}\approx 3000$~K for $A=0$, to $T_{\rm eq}\lsim 1500$~K for $A=0.95$.  The ``lava-ocean model'' would predict a relatively high albedo, in which a combination of low absorption and efficient backscatter of incident light by the melted materials could explain the high reflectivity (Rouan et al.~2011). A high albedo has also been inferred for Kepler-10b (Batalha et al.~2011).

These considerations show that the occultation and illumination signals are compatible with the planetary interpretation. Both energy considerations and the observed illumination curve amplitude suggest that the nightside is darker than the dayside, but this conclusion depends on the unknown fraction of the occultation depth that is contributed by reflection as opposed to thermal emission.  The thermal emission coming from the nightside, relative the stellar flux, is approximately $\delta_{\rm occ} - 2A_{\rm ill} = 1.7 \pm 1.6$~ppm.  We can translate this into a constraint on the temperature of the nightside, by assuming a uniform temperature and taking into account the appropriate bolometric corrections. The result is $T_{\rm n} < 2700$~K (3$\sigma$).  The dayside could have a similar temperature if the albedo is high enough, leaving open the possibility that the surface of the planet has a nearly uniform temperature---which, in turn, would imply efficient heat redistribution from the dayside to the nightside.  Observations of the secondary eclipse in one or several different wavelengths will help distinguish between different models, allowing for a better measurement of the albedo and the day-night heat redistribution efficiency.  Further modeling of the planet would also benefit from better knowledge of $a/R_\star$ and $R_p/R_\star$, which might be obtained from more finely-sampled photometric observations.

\section{Discussion}

We have interpreted the results as the discovery of the planet Kepler-78b orbiting a late G-type star with an orbital period of 8.5 hours, the shortest period among all of the well-documented planets transiting a main-sequence star. Our spectroscopic observations and light-curve analysis of Kepler-78 are all consistent with this scenario. The lack of radial velocity variations tells us that if the planet orbits Kepler-78, then its mass cannot 
be higher than $100 \, M_\oplus$. The lack of ellipsoidal light variations confirms this with an even more stringent limit of only $8 \, M_\oplus$. The ELV limit also eliminates the alternative scenario in which the 
signal is due to a blended eclipsing binary, since the transiting object must then have a mass lower than $8 \, M_{\rm Jup}$.  Those blend scenarios involving massive planets are also disfavored by the lack of anomalous image photocenter motion.

\begin{figure}[ht]
\begin{center}
\leavevmode
\hbox{
\epsfxsize=3.45in
\epsffile{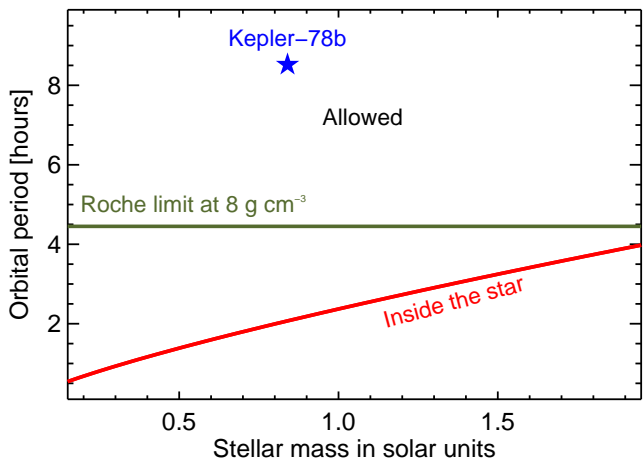}
}
\end{center}
\vspace{-0.1in}
\caption{Shortest allowed periods of rocky planets as a function of the mass of a main-sequence host star. The red curve is the limiting period at which the planet would be grazing the stellar photosphere. The green curve represents the limit on the orbital period at which the planet would be located at its Roche limit with a fiducial density of 8~g~cm$^{-3}$. We also note that for really low mass planets, even rocky material could quickly evaporate at the equilibrium temperatures implied by these close distances, making those planets inviable at distances larger than their Roche limit. This is not the case for super-Earths which may be able to survive evaporation for very long intervals at orbital periods of only a few of hours (see, e.g., Perez-Becker \& Chiang 2013).}
\label{fig:short_P}
\vspace{0.1in}
\end{figure}

With a {\it Kepler} magnitude of 11.5, large ground-based telescopes could be used to detect the transit with several repeated observations. It may also be possible to measure the mass of the planet through radial-velocity monitoring of the host star. An Earth-mass planet would induce a 1~m~s$^{-1}$ signal over a single night. The observations would require careful treatment of the starspot-induced spurious radial velocities, which should occur with amplitude $\approx$30~m~s$^{-1}$ over the 12.5-day stellar rotation period. If achieved, this would be the smallest planet with a measured mass. Currently the only planets with radius smaller than 1.8~$R_{\Earth}$ for which the mass has been measured are CoRoT-7b (L{\'e}ger et al.~2009, 2011), Kepler-10b (Batalha et al.~2011), Kepler-11b (Lissauer et al.\ 2011, 2013) and Kepler-36b (Carter et al.\ 2012), and none of these are smaller than 1.4~$R_{\Earth}$ .

The robust detections of the occultations of the planet by the star, and of the time-variable illuminated fraction of the planet as it orbits around the star, make the system important for future observational and theoretical work.  Observations with finer time sampling could better pin down the transit parameters. This in turn would clarify the equilibrium temperature of the planet's dayside, as explained in the previous section.  It is unclear at this point if the occultations would be large enough in any band to be detected with any telescope besides {\it Kepler}, but the prospect of studying the surface or atmosphere of an Earth-sized exoplanet may be attractive enough to justify a large investment of telescope time.

In addition to the questions surrounding the existence and evolution of very hot terrestrial bodies, the relative ease with which this planet was detected and vetted shows the practical importance of searches for the shortest-period planets.  We plan to document our list of short-period planet candidates found via Fourier transform analysis in the near future, which should provide other good targets for follow-up and shed some light on their occurrence rate and general characteristics.  As a final note we point out that rocky planets could well exist with even shorter periods than we have found for Kepler-78b. We show in Fig.~\ref{fig:short_P} that it is at least sensible to search for planet periods down to $\sim$4 hours (see Rappaport et al., submitted).

\acknowledgements We are indebted to Allyson Bieryla and Gilbert Esquerdo for their assistance with the spectroscopic follow-up observations. We also thank the referee, Nick Cowan, for a thorough critique of the manuscript. We thank Bryce Croll, Brice Demory, Amaury Triaud and Kevin Schlaufman for helpful discussions about this object, and of course we are grateful to the entire {\it Kepler} team for providing a wonderful tool for discovery.  R.S.O.\ and J.N.W.\ acknowledge NASA support through the Kepler Participating Scientist Program. D.W.L.\ acknowledges partial support for the spectroscopic work from the Kepler mission under NASA Cooperative Agreement NNX13AB58A with the Smithsonian Astrophysical Observatory. This research has made use of the NASA Exoplanet Archive, which is operated by the California Institute of Technology, under contract with the National Aeronautics and Space Administration under the Exoplanet Exploration Program. The data presented in this article were obtained from the Mikulski Archive for Space Telescopes (MAST). STScI is operated by the Association of Universities for Research in Astronomy, Inc., under NASA contract NAS5-26555. Support for MAST for non-HST data is provided by the NASA Office of Space Science via grant NNX09AF08G and by other grants and contracts.

\end{document}